\pdfoutput=1

\documentclass[sigplan,screen]{acmart}
\copyrightyear{2024}
\acmYear{2024}
\setcopyright{rightsretained}
\acmConference[ASPLOS '24]{29th ACM International Conference on Architectural Support for Programming Languages and Operating Systems, Volume 4}{April 27-May 1, 2024}{La Jolla, CA, USA}
\acmBooktitle{29th ACM International Conference on Architectural Support for Programming Languages and Operating Systems, Volume 4 (ASPLOS '24), April 27-May 1, 2024, La Jolla, CA, USA}
\acmDOI{10.1145/3622781.3674181}
\acmISBN{979-8-4007-0391-1/24/04}

\usepackage[]{hyperref}
\usepackage{algorithmic}
\usepackage{graphicx}
\usepackage{textcomp}
\usepackage{xcolor}
\usepackage{multirow}
\usepackage{color,soul}

\title{Litmus: Fair Pricing for Serverless Computing}

\author{Qi Pei}
\affiliation{%
    \institution{Binghamton University}
    \city{Binghamton}
    \state{New York}
    \country{USA}}
\email{pqi1@binghamton.edu}

\author{Yipeng Wang}
\affiliation{%
   \institution{Intel Labs}
   \city{Portland}
   \state{Oregon}
   \country{USA}}
\email{yipeng1.wang@intel.com}

\author{Seunghee Shin}
\affiliation{%
   \institution{Binghamton University}
   \city{Binghamton}
   \state{New York}
   \country{USA}}
\email{sshin@binghamton.edu}

\begin{document}



\begin{abstract}

Serverless computing has emerged as a market-dominant paradigm in modern cloud computing, benefiting both cloud providers and tenants. While service providers can optimize their machine utilization, tenants only need to pay for the resources they use. To maximize resource utilization, these serverless systems co-run numerous short-lived functions, bearing frequent system condition shifts. When the system gets overcrowded, a tenant's function may suffer from disturbing slowdowns. Ironically, tenants also incur higher costs during these slowdowns, as commercial serverless platforms determine costs proportional to their execution times.

This paper argues that cloud providers should compensate tenants for losses incurred when the server is over-provisioned. However, estimating tenants' losses is challenging without pre-profiled information about their functions. Prior studies have indicated that assessing tenant losses leads to heavy overheads. As a solution, this paper introduces a new pricing model that offers discounts based on the machine's state while presuming the tenant's loss under that state. To monitor the machine state accurately, Litmus pricing frequently conducts Litmus tests, an effective and lightweight solution for measuring system congestion. Our experiments show that Litmus pricing can accurately gauge the impact of system congestion and offer nearly ideal prices, with only a 0.2\% price difference on average, in a heavily congested system.

\end{abstract}

\maketitle 



\section{Introduction}
\label{sec:introduction}

Serverless computing is growing fast, satisfying today's high demands for better programmability. It allows tenants to develop their programs as stateless functions in high-level languages such as Python, Go, and JavaScript without concerns with low-level resource management and task scheduling~\cite{serverless2,serverless1,schall2022lukewarm,schall2023warming,mohammad2020SerInWild}. This advantage attracts cloud users, helping them to deploy their programs quickly with minimal effort. Recent reports note that over 50-70\% of cloud users have adopted serverless computing for their tasks~\cite{serverless1}.

Serverless computing also benefits tenants financially. Unlike traditional cloud platforms that require tenants to purchase the hardware resources needed to deploy their servers regardless of actual usage~\cite{amazonEC2Pricing}, serverless computing claims a pay-as-you-go pricing model, where tenants only pay for the resources consumed during the execution of their functions, facilitating significant cost savings. Meanwhile, this pricing requires service providers to accurately monitor and estimate each user's fine-grained resource usage~\cite{microsoftAzurePricing,awsLambdaPricing,googleCloudFunctionPricing}. 

However, today's high-performance systems maximize resource efficiency by executing multiple applications simultaneously and allowing resources to be shared~\cite{zhao2021understanding}. Such a shared serverless system raises two major challenges. First, tracking a tenant's resource usage becomes challenging, as the system can only monitor a tenant's use of exclusively dedicated resources while the system is shared. Second, and more importantly, co-running applications compete to grab more resources, affecting each other's progress~\cite{jin2015hardware,kannan2019caliper,vicent2017AppClustering,lavanya2015AppSlowdown,zhao2021understanding}. This interference not only complicates tracking resource usage but also leads to tenants being charged unfairly due to delayed execution times~\cite{alex2013ScFairPrice,goltzsche2019acctee,jin2015hardware,tiwari2013enabling}.

A serverless platform should provide a user with an isolated runtime environment. Otherwise, we argue that the platform provider should compensate a tenant's loss as a discount when the isolation is not achievable. Thus, we define a charged fee as fair when it reflects the tenant's slowdowns caused by resource sharing. The following two methods can be used to offer a fair price in a multi-tenant environment.

First, strict resource partitioning can divide shared resources into multiple partitions, each assigned exclusively to a user. This means the user only pays for the allocated resources and is free from others' interference~\cite{kim2004Partition,vicent2017AppClustering,roy2021satori,park2019copart}. However, strict partitioning is known to cancel opportunities to utilize resources efficiently~\cite{marangoz2021designing,wang2016simultaneous,el2018kpart,wang2008adaptive}. Also, a tenant's application may suffer unacceptable slowdowns when insufficient resources are allocated~\cite{roy2021satori,park2019copart}. Such a delay directly appears as an increased price. Consequently, resource partitioning inherently requires users to deduce the optimal amount of resources they need when requested. 

Alternatively, a discount-based scheme estimates a user's price relative to the slowdown experienced, discounting from the price when running the task without interference, i.e., running the task alone on the machine~\cite{alex2013ScFairPrice,tiwari2013enabling}. Ideally, this scheme benefits both cloud providers and tenants compared to partitioning. Unlike partitioning, it allows tenants to share resources, enabling providers to maximize hardware utilization and generate more revenue with the given machines. Meanwhile, users get reasonable discounts proportional to the slowdowns they experience due to co-running tasks, preventing them from overpaying for lowered service quality.

While the alternative scheme looks superior, it poses a new challenge: measuring a function's slowdown. Although the platform can measure its execution time, determining the slowdown requires knowing its baseline performance, which can be obtained when it runs without interference. However, such profiling is often prohibited due to security concerns, particularly when the function processes confidential information. Additionally, the function may behave differently, with varying inputs each time it runs. That is to say, online profiling is necessary. Unfortunately, this requires expensive runtime sampling, as presented in POPPA~\cite{alex2013ScFairPrice,tiwari2013enabling}. When assessing baseline performance, all co-running tasks must be stalled. Hundreds to thousands of concurrent short-lived functions make this approach indeed impractical, as frequently sampling the baseline performance of such numerous functions is unrealistic. Thus, a new, practical solution that can be adopted in serverless platforms is needed.



We highlight that both partitioning-based and sampling-based approaches complicate the problem, either sacrificing the machine's efficiency or appearing impractical for serverless computing. Approaches in serverless computing must be lightweight, frequent, and sensitive enough to deal with quickly changing environments with numerous concurrent functions. This paper proposes Litmus pricing, a method that efficiently estimates a tenant's price with a discount proportional to the level of system congestion without burdening system performance. 

During each function's execution, Litmus pricing conducts a Litmus test, which measures the level of system congestion without adding extra overhead, using the function's startup process. Litmus test is based on the observation that serverless functions perform largely identical operations during their startup, typically involving significant memory accesses. Additionally, Litmus pricing splits the hardware resources into "private" and "shared" categories, proposing different charging ratios for each. This approach further enhances the accuracy of the discount calculation.

This study investigates Litmus pricing across diverse environments with varying levels of system congestion. We confirm that crowded environments pose complex challenges. Despite this, Litmus pricing accurately estimates a tenant's cost, deviating by only 0.2\% on average from the ideal price that discounts tenants proportional to slowdowns.

\section{Background}
\label{sec:background}

\textbf{Serverless Computing:} Serverless computing is a new cloud execution model that allows cloud tenants to focus on developing front-end applications while leaving back-end implementation managed by service providers~\cite{serverless1, schall2022lukewarm}. This model benefits tenants by easing their burdens to secure and maintain the hardware resources and allowing them to write their code (functions) in high-level languages (e.g., Python or Node-js) as event handlers~\cite{serverless1,mohammad2020SerInWild,schleier2021serverless}. These functions are later invoked upon associated events. The serverless platform ensures security by executing the tenant's function in a sandbox, such as a container or a virtual machine~\cite{awsLambdaPricing,googleCloudFunctionPricing,microsoftAzurePricing,zhang2021faster}.

Each tenant's function is stateless and expected to be short-lived~\cite{serverless1,serverless2,eismann2020review,mohammad2020SerInWild,jia2021nightcore}. Using these short-lived functions, tenants can easily scale by invoking many functions without provisioning the needed resources for themselves, relying instead on the serverless platform~\cite{ska20ISCABabelFish,mohammad2020SerInWild}. In serverless computing, numerous short-lived serverless workloads run together on a single machine, making the execution environment more dynamic than the traditional cloud~\cite{ustiugov2021benchmarking,schall2022lukewarm,tariq2020sequoia}. Tenants need to trust the service provider to provision enough necessary resources, expecting the quality of service defined in the service level agreements (SLA) that the provider and tenants mutually agreed upon~\cite{awsLambdaPricing,googleCloudFunctionPricing,microsoftAzurePricing}.

In traditional cloud computing, tenants purchase hardware resources for a contracted term to support their applications, which are typically large servers. Thus, tenants must pay for the agreed period regardless of actual usage. In contrast, serverless workloads are generally small and launched on-demand as short-lived event handlers. This difference leads to a unique pricing policy known as pay-as-you-go. Under this model, tenants are charged only for the resources consumed during execution, typically calculated as the product of execution time and assigned memory capacity~\cite{awsLambdaPricing,googleCloudFunctionPricing,microsoftAzurePricing}.

\textbf{Language runtime:} Python and Node.js stand out as the most popular runtimes in AWS Lambda, with Python used in 58\% of all Lambda functions and Node.js in 31\%~\cite{serverless3}. Over 90\% of organizations choose Python and Node.js for their AWS services. Other popular runtimes include Java, Go, .NET, and Ruby~\cite{serverless1,serverless2}. These high-level languages offer significant abstraction from machine code with good programmability and portability, which in turn improves programmer productivity. Consequently, they are widely adopted by serverless users. However, high-level languages typically rely on runtime interpretation, leading to slower program startup due to extra launching steps. For example, when an application is launched in Python, the interpreter is first prepared. It then parses command-line arguments and imports necessary modules. Afterward, the codes are compiled and executed. Upon completion, the Python program terminates, and all allocated resources are released~\cite{pythonUsage}.

\section{Methodology}
\label{sec:method}


This section outlines our test setup for evaluating serverless functions in this study.

\textbf{Hardware Infrastructure:} All experiments are conducted on a dual-socket server platform equipped with two Intel Xeon Gold 5218 processors based on the Cascade Lake architecture. Each CPU provides 16 cores that support simultaneous multithreading (SMT) with a maximum frequency of 3.9 GHz. Every core features 32KB of L1 instruction/data caches and a 1MB L2 cache. All cores in each socket share a 22MB L3 cache. The two sockets collectively have access to 384GB of main memory. The server operates on Ubuntu Server OS, version 22.04 LTS, with kernel version 5.15.0. We use Linux Perf~\cite{perfWiki} to gather performance counters.

The CPU frequency is a critical factor influencing both performance and energy consumption, significantly affecting a function's execution time and energy usage. A CPU's frequency can be adjusted through software or hardware based on power and thermal budget, optimizing for either energy efficiency or performance. However, varying frequencies can cause instability in system performance and energy usage, complicating system management. For this reason, commercial systems like Google Cloud offer only one fixed frequency for their vCPUs~\cite{googleCloudFunctionPricing}. Accordingly, we set our CPUs' frequency at 2.8GHz. If we do not fix the frequency through software, Intel's Turbo technology occasionally adjusts it, but it mostly remains at 2.8 GHz during our tests. We will provide more details on this in Section~\ref{sec:sensitivity}.

\begin{table}[h]
    \begin{center}
        \caption{Serverless Benchmarks \& Language Runtimes (py, nj, go)}
        \label{tb:benchmarks}
        \begin{tabular}{|l l||l l|} \hline
            Function & Abbr. & Function & Abbr. \\ \hline \hline
            \multicolumn{2}{|c||}{\textbf{SeBs~\cite{copik2021sebs}}}  &  \multicolumn{2}{|c|}{\textbf{Function Bench~\cite{kim2019functionbench}}}\\ \hline
            Dyn HTML & dyn-py & Chameleon & chame-py \\ \hline
            Thumbnail & thum-py* & FloatOp & float-py \\ \hline
            Compression & compre-py & Gzip & gzip-py* \\ \hline
            Image Recogn & recogn-py & RandDisk & randDisk-py* \\ \hline
            Graph Rank & pager-py & SequenDisk & seqDisk-py \\ \hline
            Graph Mst & mst-py & \multicolumn{2}{c|}{\textbf{Online Boutique~\cite{googleMicroserviceDemo}}} \\ \hline
            Graph Bfs & bfs-py* & Currency & cur-nj* \\ \hline
            DNA Visual & visual-py* & Payment & pay-nj \\ \hline
            \multicolumn{2}{|c||}{\textbf{Hotel Reservation~\cite{gan2019open}}} & \multicolumn{2}{c|}{\textbf{Other~\cite{awsLambdaApiGateWay,copik2021sebs}}}  \\ \hline
            Geo & geo-go & Authen & auth-py*/nj/go \\ \hline
            Profile & profile-go* & Fibonacci & fib-py*/nj*/go* \\ \hline
            Rate & rate-go & AES & aes-py/nj*/go* \\ \hline
        \end{tabular}
    \end{center}
\end{table}

\textbf{Workload:} As listed in Table~\ref{tb:benchmarks}, we selected 27 distinct serverless functions from various benchmark suites, including AWS authentication serverless functions~\cite{awsLambdaApiGateWay}, the Hotel Reservation from DeathStarBench~\cite{gan2019open}, Google's Online Boutique application~\cite{googleMicroserviceDemo}, FunctionBench~\cite{kim2019functionbench}, and the SeBS serverless benchmark suite~\cite{copik2021sebs}. The functions are implemented using three languages: Python, Nodejs, and Go. Notably, the three functions, Authen, Fibonacci, and AES, are implemented in all three languages, creating three separate test cases per function. 
We installed Python 3.10.6, Nodejs v12.22.9, and Go 1.19.2 to support them. Additionally, we selected 13 benchmarks (* marked) in Table~\ref{tb:benchmarks} as our reference applications, which are explained later.



\begin{figure}[h]
\centering
\includegraphics[width=1\columnwidth]{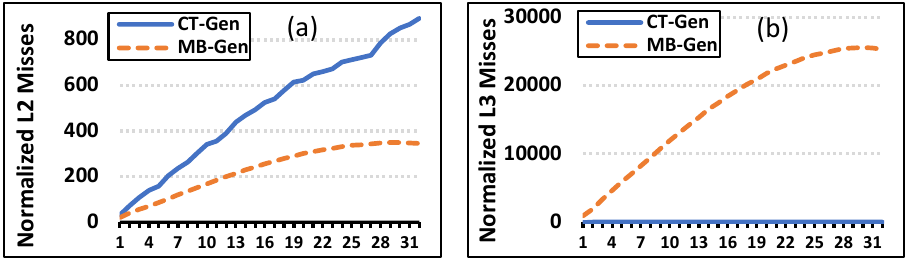}\\
\caption{(a) L2 misses and (b) L3 misses of traffic generators, both normalized with the average L2 and L3 misses of serverless applications listed in Table~\ref{tb:benchmarks}}
\label{fig:stressing_specs}
\end{figure}

\textbf{Traffic Generator:} We categorize the CPU's shared resources into two groups: those before the L3 caches and those after. To stress each group independently, we developed two separate traffic generators designed to generate traffic and congestion within these resources. The first generator, CT-Gen, exerts pressure from the core up to the L3 cache by creating substantial memory accesses that mostly miss the L2 cache but hit the L3 cache. Another generator, MB-Gen, induces massive L3 misses, which stress the resources beyond the L3 cache, primarily targeting off-chip memory bandwidth and L3 cache spaces.

Both traffic generators are multi-threaded and govern the amount of traffic by adjusting the number of threads they spawn. Each thread is pinned to a specific core to avoid conflicts between threads, allowing us to vary the stress level on our 32-core processors from level 1 to 31. Figure~\ref{fig:stressing_specs} illustrates each generator's characteristics. In Figure~\ref{fig:stressing_specs}(a), with each increase in thread count, CT-Gen generates substantial L2 cache misses. These L2 misses by CT-Gen are ended with L3 cache hits, as shown in Figure~\ref{fig:stressing_specs}(b). In contrast, MB-Gen produces significant L2 and L3 cache misses. Interestingly, MB-Gen's L2 misses are fewer than those of CT-Gen because MB-Gen is hindered by its L3 misses, suffering from a self-imposed performance bottleneck.

\section{The Need for Fair Pricing}
\label{sec:motivation}

Fairness in cloud computing, especially when multiple tenants share hardware resources, is addressed through various approaches. One standard method is the Service Level Agreement (SLA), which specifies the level of service the cloud provider must deliver and the tenants' expectations. However, the SLA outlines the minimum quality of service the providers should deliver rather than solving fairness issues. Despite serverless computing's emphasis on a pay-as-you-go pricing model, unfairness among tenants still exists.

Unfortunately, modern systems are designed to maximize hardware efficiency by running multiple applications concurrently. This results in conflicts over shared resources, affecting each function's execution time. Ironically, when service providers optimize for profit by packing more functions into a single machine, the execution time for each function increases, resulting in higher costs for the same workload. In serverless computing, tenants have no control or visibility over how their software is executed, leading to potential inefficiencies and higher costs for the sake of convenience.

In serverless computing, although each function runs exclusively in a virtual sandbox with private resources, it still shares certain resources like caches and memory bandwidth with other functions. Resource partitioning, a traditional solution that eliminates interference by dedicating resources exclusively to a tenant, cannot avoid inefficient resource utilization. It is also against the serverless philosophy, demanding the tenants govern their hardware resources. 

\begin{figure}[h]
\centering
\includegraphics[width=1\columnwidth]{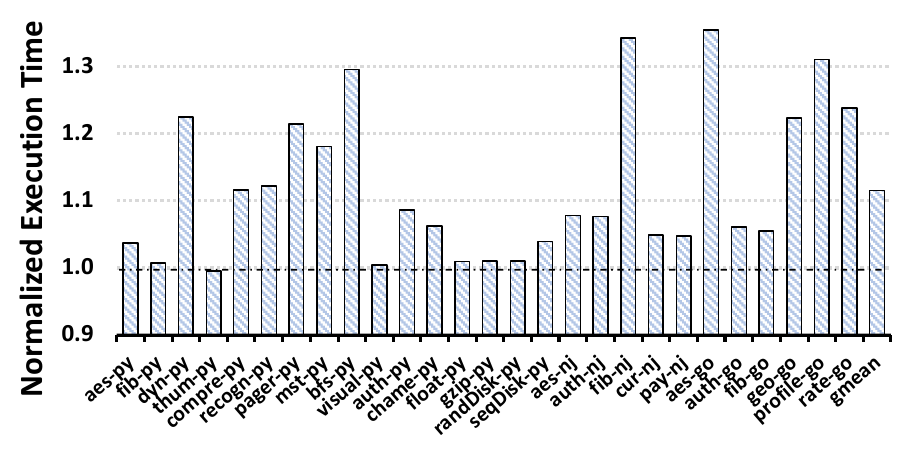}\\
\caption{Execution time of applications that run with 26 others, normalized to the execution time when running alone}
\label{fig:normalizedCPIRunWith26Rev}
\end{figure}

Figure~\ref{fig:normalizedCPIRunWith26Rev} shows how much a function can be slowed when co-running with 26 other workloads. The workloads are randomly selected from the benchmarks listed in Table~\ref{tb:benchmarks}. Whenever a function finishes, a new randomly-selected function is launched to maintain a total of 26 co-running functions. The figure shows a function's performance can drop by up to 35\%, with most functions significantly impacted by the co-running functions. On average, functions perform 11.5\% slower in this shared environment compared to an isolated one, resulting in users being charged 11.5\% more. We highlight that this scenario involves only 26 co-running functions, which is tiny compared to heavily crowded commercial serverless platforms that run hundreds to thousands of functions simultaneously~\cite {ustiugov2021benchmarking,schall2022lukewarm,tariq2020sequoia}.

Instead of guaranteeing invariant performance, POPPA~\cite{alex2013ScFairPrice,tiwari2013enabling} offers price discounts to compensate for slowdowns. To provide these discounts, POPPA measures a task's slowdown due to interference through sampling and determines the discount amount based on the observed slowdowns. Unlike partitioning, this scheme requires the service provider to carry out all responsibilities related to contentions from resource sharing, ensuring tenants pay a consistent price for their usage. Any unexpected slowdowns that a tenant experiences are compensated with a discount. However, sampling is an expensive operation that requires stalling all co-running processes during the sampling. Moreover, accurate measurement necessitates frequent sampling, especially in rapidly changing environments, making the approach impractical for dynamic environments like serverless platforms.

Overall, tenants need a new pricing model that considers more than just execution time. Prior approaches have limitations, excessively burdening cloud participants and being unsuitable for the many short-lived functions typical in serverless computing. This motivates us to propose a new solution: Litmus pricing, which discounts prices based on the system congestion while remaining lightweight and imposing no additional overhead.

\section{Litmus Pricing}
\subsection{Overview}

The estimated prices of serverless functions should reflect the system's congestion state. The presence of numerous small transient functions within the serverless system can rapidly alter its congestion state, causing sudden congestion or swift resolution. Nevertheless, such transient traffic jams can significantly impact the performance of short-lived functions. Thus, assessing the congestion must be lightweight, quick, and timely. However, a prior approach, POPPA~\cite{tiwari2013enabling,alex2013ScFairPrice}, relies on sampling, which incurs considerable performance costs, thereby limiting its widespread adoption. Alternatively, this study introduces Litmus pricing, which promptly gauges system congestion at the onset of each function and reflects the estimated congestion level into pricing the function.

Like the prior approach, Litmus pricing suggests that service providers should offer discounts when a tenant's functions experience interference, thus compensating for the performance loss. Litmus pricing determines the discount rate based on the system's congestion level. Evaluating congestion also assists providers in estimating remaining resources and making informed decisions regarding job scheduling.

However, Litmus pricing does not quantify the slowdown of each individual function; the discount amount is solely determined by the severity of the measured congestion. Even though different functions face the same congestion level, their respective impacts can vary, making precise estimation of each function's performance loss a challenging task. This complexity necessitated the expensive sampling relied upon by the previous approach. Instead, Litmus pricing circumvents such overheads and lightens the service provider's burden solely to measure ongoing congestion levels.

Litmus pricing captures the dynamic significance of shared resources under varying system conditions by offering uniform discounts to all functions under the same system congestion, regardless of their individual slowdowns. Embedded within this policy is the distribution of pricing responsibility to both parties. During periods of system congestion, shared resources become more valuable and meaningfully influence a function's performance. We observed that functions experiencing larger slowdowns tend to demand more of the crowded resources, thereby exacerbating adverse overcrowded conditions. However, Litmus pricing does not offer additional compensation for the larger loss experienced by functions that use shared resources more heavily than others.

As a result, Litmus inherently incentivizes users who use shared resources sparingly. During periods of overcrowding, functions with minimal needs for shared resources still enjoy discounts but with only slight slowdowns. This implies that users may take advantage of this policy by minimizing their reliance on shared resources, receiving substantial discounts despite only experiencing minor slowdowns. This trade-off is considered acceptable because reduced shared resource usage by users helps mitigate congestion, thereby benefiting the overall system and other users.

\subsection{Pricing Model}
\label{pricing model}

Developing Litmus pricing raises two challenges. First, it needs a new model to estimate a function's price. Second, it needs a way to measure the system's congestion state. Due to the rapidly changing environment in serverless computing, assessing congestion must be lightweight and timely. Commercial serverless platforms employ a uniform pricing model regardless of resource congestion dynamics. Consequently, tenants end up paying more for prolonged execution time when using over-crowded resources. Rather than simply imposing the price by multiplying the main memory usage and a function's execution time, Litmus pricing determines the price as a sum of two distinct pricing components as follows.

\begin{equation} 
P = P_{private} + P_{shared}
\end{equation}

In this equation, \(P\) denotes the total price tenants must pay to execute their functions. Litmus pricing divides this cost into two pricing components, each associated with a different resource type. \(P_{private}\) is the cost for private resources, which are entirely assigned to a single tenant. \(P_{shared}\) is the cost for shared resources. Each pricing component is measured proportional to the time users occupy the respective resources. However, if contention disturbs a process's execution in the shared resources, the occupied time will be extended. This delay unfairly increases the tenant's cost.

\begin{figure}[h]
\centering
\includegraphics[width=1\columnwidth]{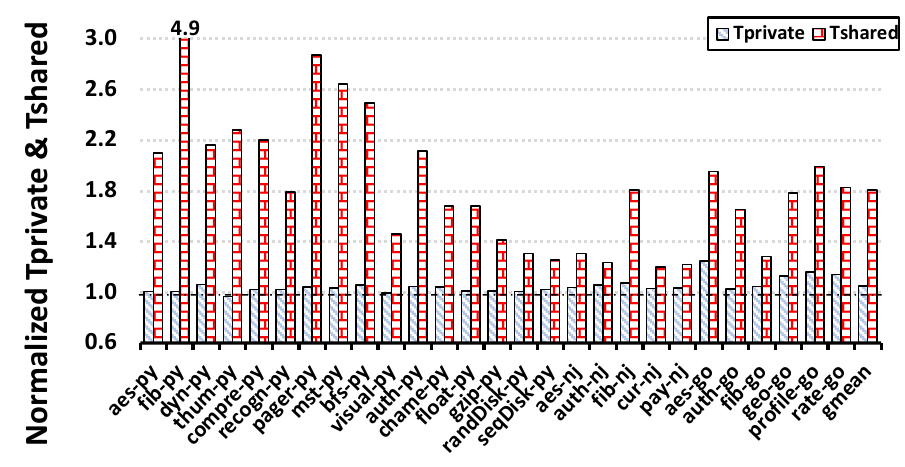}\\
\caption{\(T_{private}\) and \(T_{shared}\) of applications that run with 26 others, normalized to those when running alone}
\vspace{-0.5em}
\label{fig:slowdownRunwith26Rev}
\end{figure}


We observed that system congestion influences the occupied time for each resource type differently. The time to use private resources is only slightly affected, with minor changes, even under heavy congestion. In contrast, the time spent on shared resources is broadly affected, showing significant differences. Thus, we separately measure the delays for the two resource types and apply different charging rules for their use. 

Using Perf~\cite{perfWiki}, we split a function's execution time into two slices: \(T_{private}\), the time spent on private resources, and \(T_{shared}\), the time spent on shared resources. Litmus pricing collects \(T_{shared}\) by reading the cycles stalled due to L2 cache misses. \(T_{private}\) is calculated as the total execution cycles to complete the application minus \(T_{shared}\). Figure~\ref{fig:slowdownRunwith26Rev} shows the impact of the contentions on each time slice, where \(T_{shared}\) per instruction and \(T_{private}\) per instruction are normalized to the same tests when running alone. The figure shows that \(T_{shared}\) is significantly affected by system congestion compared to \(T_{private}\). On average, \(T_{shared}\) increases by 181\%, with a maximum increase of 488\%. Meanwhile, \(T_{private}\) increases by only 4\%, with minor variance between functions.

This figure highlights two key factors. First, congestion in shared resources has a negligible impact on the performance of private resources. Accordingly, when the system gets congested, the extent of a function's reliance on shared resources determines its performance. If \(T_{shared}\) constitutes a minor portion of a function's execution, even severe congestion will have minimal impact on its overall execution. Second, hardware Performance Monitoring Units (PMUs) in modern CPUs and profiling tools like Linux Perf are widely available and can be utilized to differentiate between the use of shared and private resources. Throughout this paper, we will leverage these tools to develop Litmus pricing.

\begin{figure}[h]
\centering
\vspace{-0.5em}
\includegraphics[width=1\columnwidth]{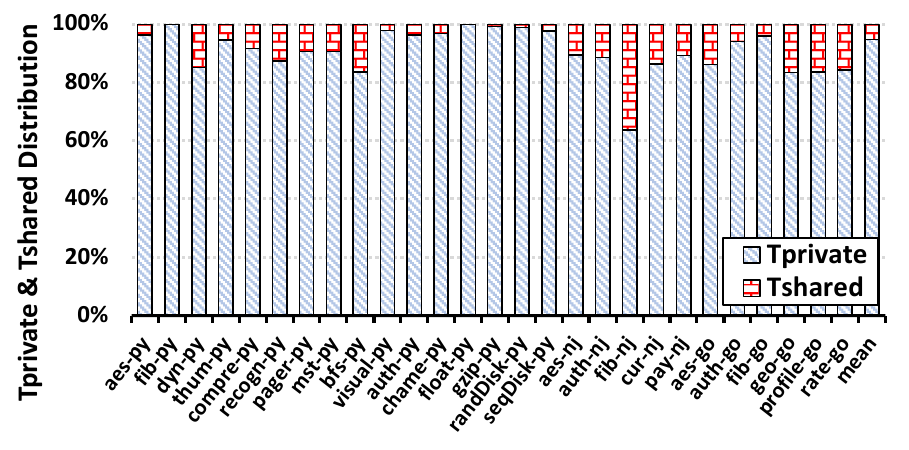}\\
\caption{Execution time distribution of \(T_{private}\) and \(T_{shared}\)}
\vspace{-0.5em}
\label{fig:TpriTshardDistriRev}
\end{figure}

Figure~\ref{fig:TpriTshardDistriRev} shows the distribution of \(T_{private}\) and \(T_{shared}\) during a function's execution. Compute-intensive workloads (e.g., fib-py, float-py) exhibit a substantial \(T_{private}\) portion, accounting for up to 99.96\% of the execution time. However, this portion decreases when running memory-intensive workloads such as fib-nj. This distribution reflects how other co-running functions can influence its execution. For example, although running float-py in a crowded system may increase its \(T_{shared}\), the total execution time, $T_{private}+T_{shared}$, will undergo only a minor change.

After collecting the time slices spent on each resource, Litmus calculates each pricing component by multiplying two different charging rates with corresponding slices. Consequently, a new pricing equation emerges as follows:

\begin{equation} 
P = R_{private} * T_{private}  + R_{shared} * T_{shared}
\end{equation}


\(R_{private}\) and \(R_{shared}\) are the charging rates for private and shared resources, determined by the congestion level. These two charging rates are necessary because each component is affected differently, either directly or indirectly. The charging rate determines the actual discount amount, reflecting the degradation of applications' performance caused by interference from a crowded execution environment. Thus, the charging rates should be measured as follows:

\begin{equation} 
R = R_{base} * \frac{T_{solo}}{T_{congestion}}
\end{equation}

In this equation, \(R\) represents the charging rate for executions on private or shared resources. \(R_{base}\) is the base charging rate when the function runs in a congestion-free runtime environment; our tests select 1 for \(R_{base}\). \(T_{solo}\) is its non-interfered execution time when the function runs alone, and \(T_{congestion}\) is its execution time in a congested environment. We need to collect separate \(T_{solo}\) and \(T_{congestion}\) values for each pricing component. To measure \(T_{solo}\), the prior study~\cite{tiwari2013enabling,alex2013ScFairPrice} employs an expensive approach that suspends all co-running applications whenever sampling an application's non-interfered performance. However, Litmus pricing avoids such a costly approach by presuming the function's solo performance through reference workloads on the system.

Litmus pricing requires both providers and tenants to be involved in the price decision process. Providers need to establish reasonable rates for each resource usage by analyzing selected representative functions. To offer fair discounts, providers must carefully choose the functions to be analyzed, ensuring the functions accurately represent the system condition. Tenants must also know the consequences when their tasks heavily rely on shared resources. Unfortunately, their tasks may get delayed more than the provider's expectations while still receiving the expected discount rate. 


\section{Estimating prices with Litmus}
\label{sec:designDetails}

\begin{figure}[h]
\centering
\includegraphics[width=1\columnwidth]{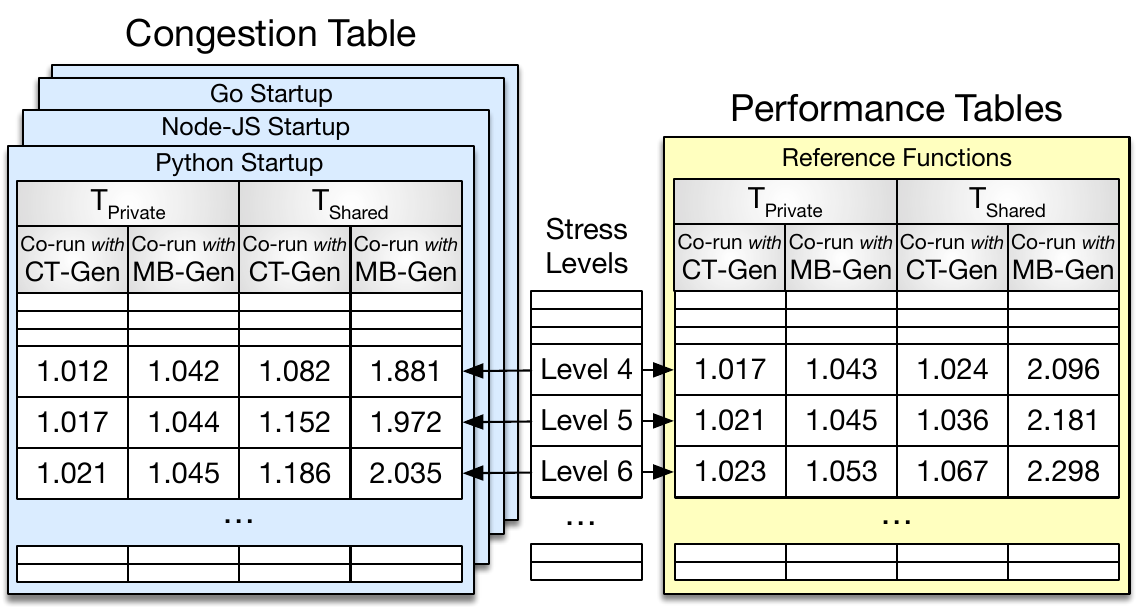}\\
\caption{Congestion and performance tables: numbers in both tables indicate the slowdowns of startup codes and reference functions, collected with CT-Gen and MB-Gen}
\vspace{-0.7em}
\label{fig:litmust_tables}
\end{figure}

Before using Litmus pricing, service providers need to complete the congestion and performance tables, shown in Figure~\ref{fig:litmust_tables}. The congestion table lists the slowdowns during the startup phases, whereas the performance table lists the slowdowns of reference functions. Litmus pricing needs three steps to determine the final price with a discount. First, providers assess system congestion at multiple levels and record the results in the congestion table. Second, providers analyze the impact of congestion on reference functions and fill out the performance table. Lastly, providers determine a tenant's price based on the data from both tables.

\textbf{Step 1, Defining Congestion Levels:} We manage the system's congestion levels using two traffic generators that create a range of congestion states by stressing shared resources differently. The traffic generator is multi-threaded, creating traffic in shared resources to incur congestion, with each thread running on a separate core. We rank the stress levels from 1 to 32 by changing the number of active threads running on distinct cores. Entries from top to bottom in the congestion table indicate increasing stress levels.



Measuring the amount of system congestion and its impact on performance is a crucial challenge. This challenge becomes particularly pronounced when trying to quantify the impact of the congestion on a running application without pre-profiled information about the application. The fast-changing system states of a serverless platform, with numerous short-lived functions, further complicate the assessment, necessitating quick and frequent performance examinations.

Litmus pricing includes a novel method, named a \textit{Litmus test}, that examines the system's state quickly without imposing additional overheads. This method leverages our observation that high-level languages like Python and Node.js, commonly used in serverless computing, have lengthy startup phases to accommodate their runtime environments. Since the inception of serverless computing, these startup delays have been identified as a major source of latency issues~\cite{ismail2018quantitative}. Litmus tests exploit these startups to read the system's state, utilizing their fixed routines with consistent operations. Our tests have validated that tested functions exhibit similar performance characteristics during startup phases.


\begin{figure}[h]
\centering
\includegraphics[width=1\columnwidth]{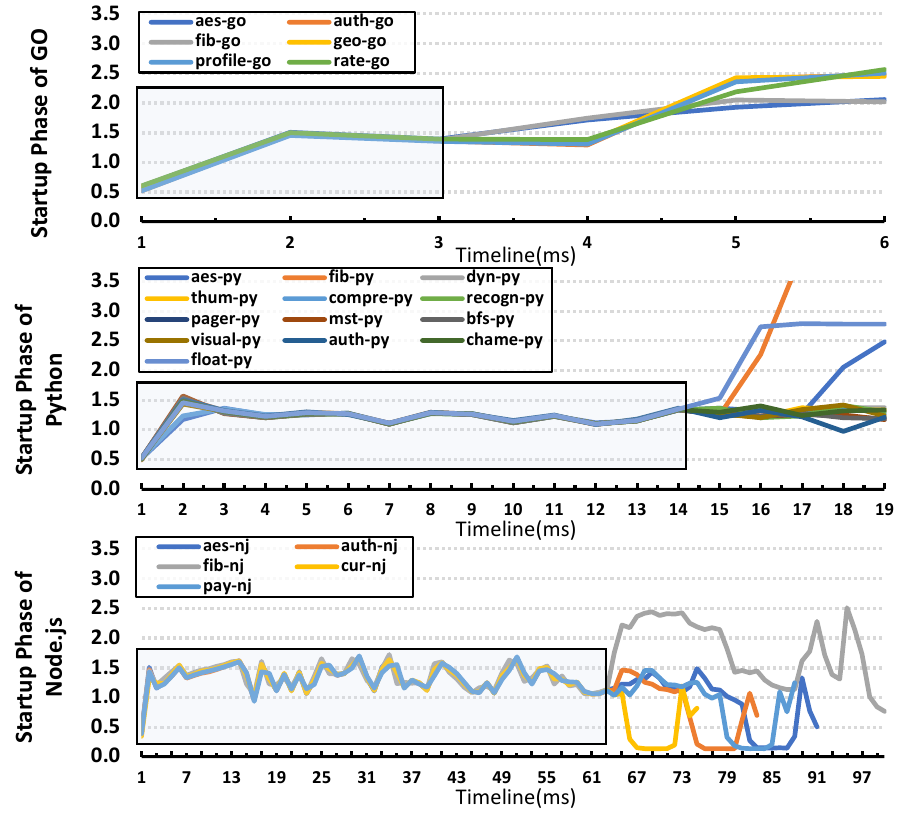}\\
\caption{IPC of serverless applications based on languages (Go, Python, and NodeJS) during their startup phase}
\label{fig:startupIPC_grouped}
\end{figure}

Figure~\ref{fig:startupIPC_grouped} depicts a function's fluctuation in IPC (Instructions Per Cycle) during its startup phases. The x-axis represents the time elapsed since the beginning, with each tick indicating one millisecond. The y-axis shows the function's IPC measured for a millisecond at corresponding time intervals. We picked functions with different languages and characteristics from those listed in Table~\ref{tb:benchmarks}. These functions were categorized into three groups based on language runtime, and the IPC of each group is presented in separate figures. As shown in these figures, functions in the same figure show notable similarities. This implies that functions written in the same language have nearly identical startups.

\begin{figure}[h]
\centering
\includegraphics[width=1\columnwidth]{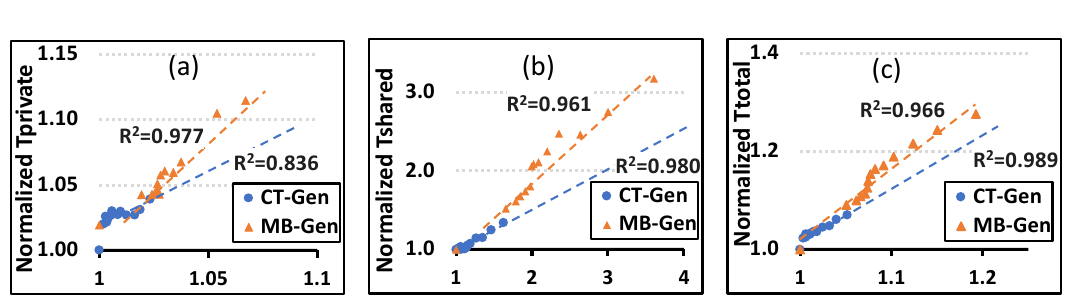}\\
\caption{The correlation between the slowdown of Python's startup phase and the slowdown of reference applications. From left, (a) \(T_{private}\), (b) \(T_{shared}\), and (c) \(T_{total}\), normalized to the execution time when running alone}
\label{fig:referenceStartupSlowdownRev}
\end{figure}

The startup of a runtime typically includes bursts of memory reads to load images and libraries, which can be used to probe the traffic in shared resources. By conducting a Litmus test before executing a tenant's function, providers can assess the system state. Meanwhile, providers also need to study how startup phases perform under different system states. They complete the congestion tables by conducting the Litmus tests with startups written in three languages while varying congestion levels. This information from the congestion table is then referenced when constructing the congestion model in Figure~\ref{fig:referenceStartupSlowdownRev}.

\begin{figure}[t]
\centering
\includegraphics[width=1\columnwidth]{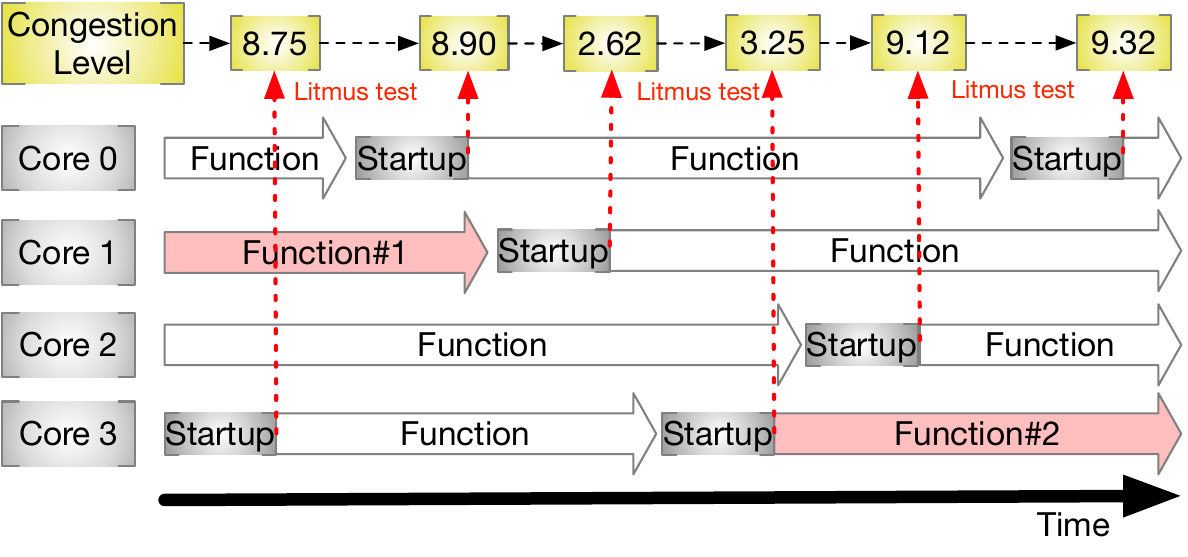}\\
\caption{Progress of serverless applications on four cores with Litmus tests (dotted arrows) over time}
\label{fig:Litmus_tests}
\end{figure}

Figure~\ref{fig:Litmus_tests} shows the progress of serverless functions with Litmus tests, which examine the level of system congestion during each function's startup. Many concurrent functions in serverless offer frequent opportunities for testing. Function \#1, which is memory-intensive, causes congestion levels to rise above 8 during its execution. A Litmus test in Core 3 detects this congestion. After Function \#1 is complete, a subsequent Litmus test indicates a non-congested system with a congestion level below 3. However, during another resource-intensive function, Function \#2, the Litmus test identifies new system congestion with a level above 9.  

\textbf{Step 2, Measuring the Impact of Congestion:} Discerning system congestion through the Litmus test solves only one of the two challenges; the other is determining its impact on a tenant's function. Different tenants' functions could experience varying degrees of slowdown under the same congestion level. To address this, we propose using pre-selected reference functions to analyze these impacts, thereby avoiding the cost of runtime profiling. For this, providers need to carefully select reference functions that represent tenants' functions operating on the systems, as these selections influence the overall accuracy of congestion measurement. Later, in Section~\ref{sec:evaluation}, our tests will show that pre-selected reference functions can effectively represent unknown user applications. For this study, the selected reference applications (* marked) are listed in  Table~\ref{tb:benchmarks}. 

\begin{figure}[h]
\centering
\includegraphics[width=1\columnwidth]{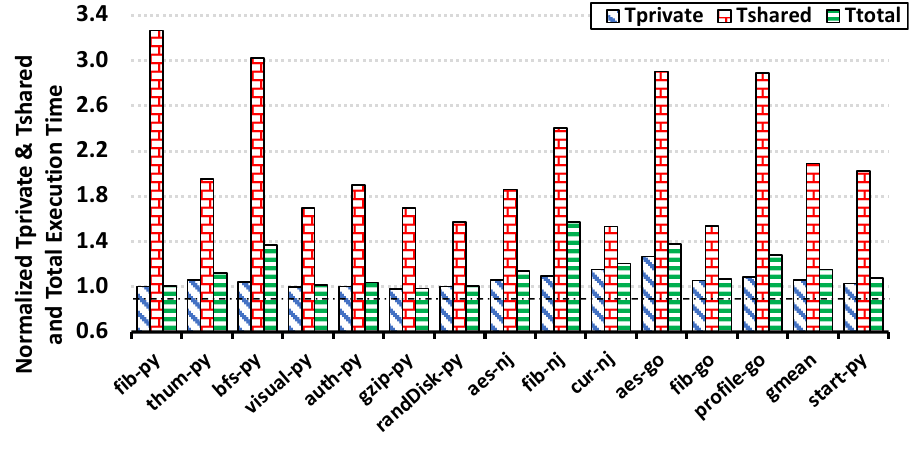}\\
\caption{\(T_{private}\), \(T_{shared}\), and \(T_{total}\) of serverless applications that run with MB-Gen at stress level 14, normalized to their execution time when they run alone}
\label{fig:refSlowdownStressing14}
\end{figure}

Once the reference functions are ready, providers analyze their slowdowns under varying congestion levels. This analysis completes the performance table shown in Figure~\ref{fig:litmust_tables}. Each entry in the table holds the geometric average of all reference functions' slowdowns at the corresponding stress level. With two traffic generators, providers can complete another set of two sub-tables for the performance table. All entries in this table are mapped 1-to-1 to entries in the congestion table. Providers can estimate a general function's slowdown at a given congestion level using this one-to-one mapping.

Figure~\ref{fig:refSlowdownStressing14} shows an example of the reference functions' slowdowns when stressing the system. The X-axis shows tested functions, each depicted with three bars: the first two bars show \(T_{private}\) and \(T_{shared}\), while the last bar shows the total execution time, all normalized to their execution time when running alone without interference. The figure shows that the tested functions experience varying degrees of slowdown despite consistent stress levels maintained during the tests. The next-to-last group shows the gmean of all the slowdowns, which providers use to estimate a function's slowdown later.


\textbf{Step 3, Determining Discount Rates:} Providers can determine a function's price through the Litmus test once the congestion and performance tables are ready. Upon obtaining a Litmus test result, the current congestion level is estimated by comparing the measured slowdown to the pre-studied slowdowns in the congestion tables. Then, Litmus pricing conjectures a possible slowdown due to the estimated congestion by using the congestion level as an index to the performance table. However, while congestion levels vary continuously, the tables only contain slowdowns at discrete intervals. Therefore, providers need a model that estimates the slowdown for any given congestion level. In our tests, we employ linear regression to develop the model.

\begin{figure}[h]
\centering
\includegraphics[width=1\columnwidth]{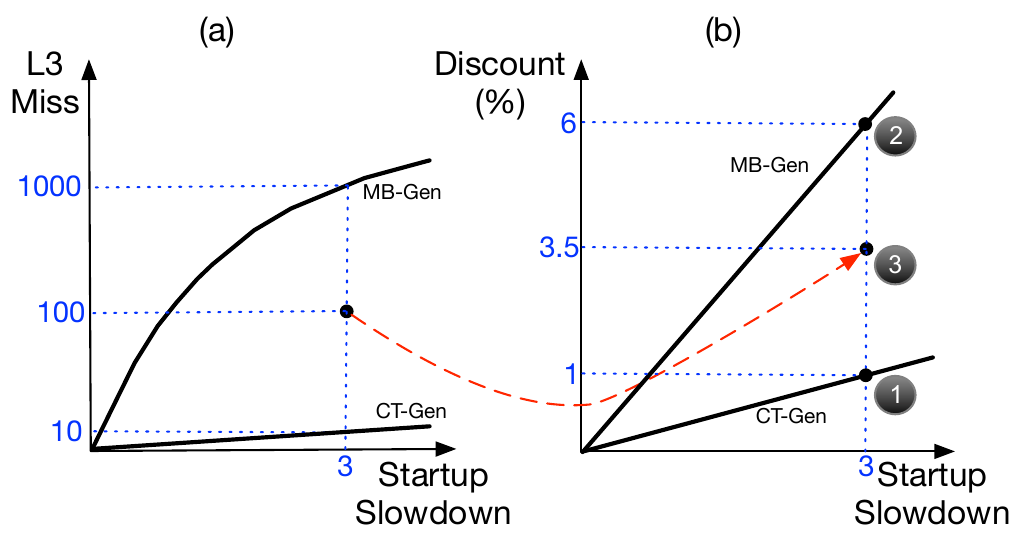}\\
\caption{Estimating a function's slowdown with logarithmic interpolation}
\label{fig:interpolation}
\end{figure}

Our experiments employ linear regression analysis to draw Figure~\ref{fig:referenceStartupSlowdownRev}, which shows the correlation between the two slowdowns. The X-axis shows the startup's slowdown, while the Y-axis shows the corresponding reference applications' slowdowns at the same congestion level. From left to right, the figure shows normalized \(T_{private}\), \(T_{shared}\), and \(T_{total}\). To extrapolate a precise congestion level, we created two sub-tables within both the congestion and performance tables using two distinct traffic generators. As a result, the figure depicts two linear regression lines (red and blue dotted lines), each associated with a different traffic generator.



These regression models return two different slowdowns. To estimate a function's slowdown, providers must determine which model reflects the current environment more accurately. The traffic generators create two extreme congestion scenarios: one before the L3 cache and one after. CT-Gen generates substantial traffic from the core to the L3 cache that filters the traffic and prevents operations from consuming memory bandwidth. Conversely, MB-Gen also generates massive traffic but consumes memory bandwidth and repeatedly evicts L3 cache blocks. Because each traffic generator represents a distinct extreme case, the actual machine state could fall somewhere between the two congestion levels created by CT-Gen and MB-Gen. However, Litmus pricing requires more than just the startup slowdown to identify the machine's state in these models. Thus, we have enabled Litmus tests to gather the system's L3 miss count as a supplementary metric that offers deeper insights into the crowdedness of shared resources.

Figure~\ref{fig:interpolation} shows our approach to using L3 misses to estimate slowdowns. We created two figures illustrating separate regression models. The left figure (a) is derived from the startup slowdowns and L3 misses with logarithmic regression, indicating L3 misses at specific congestion levels. The right figure (b) is derived from the performance table with linear regression, displaying discounts at specific congestion levels. Each figure depicts two regression lines associated with different traffic generators.

The two traffic generators set the upper and lower boundaries for L3 misses. For instance, if a Litmus test reports 10 L3 misses, close to what CT-Gen generates: \textcircled{1}This suggests congestion occurs before reaching the L3 cache, similar to CT-Gen's scenario. Thus, Litmus pricing estimates the function's discount to be around 1\%. Conversely, if the Litmus test reports 1000 L3 misses, close to MB-Gen's results: \textcircled{2}Litmus pricing estimates the function's discount rate to be approximately 6\%, indicating congestion similar to MB-Gen, where memory bandwidth is heavily consumed and L3 cache blocks are frequently evicted. When the Litmus test reports 100 L3 misses, an interpolation between these extremes is applied: \textcircled{3}Using logarithmic interpolation, Litmus pricing estimates the function's discount rate to be midway between CT-Gen's 1\% and MB-Gen's 6\%, resulting in approximately 3.5\%. Upon obtaining the discount rates, Litmus pricing computes the final price, as discussed in Section~\ref{pricing model}.

\section{Evaluation}
\label{sec:evaluation}

We split our functions into two groups: as listed in Table~\ref{tb:benchmarks}, one consists of functions to test Litmus pricing, while the other consists of chosen reference applications to derive linear regression models for Litmus pricing. The discount models plug into the Litmus pricing employed in the tests. In this study, we estimate a function's price in three ways: the price using the Litmus pricing, a commercial price that offers no discounts, and an ideal price that provides an exact discount proportional to its slowdown. 


\subsection{One Function Per Core}

We maintain consistent congestion levels across all tests by co-running each function with 26 other functions. As outlined in Section~\ref{sec:motivation}, whenever a function finishes, another is randomly launched to keep a total of 26 co-running functions. Using Perf~\cite{perfWiki}, we collect three key statistics: \(T_{private}\), \(T_{shared}\), and the machine's L3 misses during the first 45 million instructions of the Python startup. Through these metrics, Litmus pricing estimates the level of system congestion and the discount rates a function receives. Each function is executed 30 times, and we average its \(T_{private}\) and \(T_{shared}\) values. Subsequently, we calculate each function's \(P_{private}\), \(P_{shared}\), and total price (P) by summing \(P_{private}\) and \(P_{shared}\).

\begin{figure}[h]
\centering
\includegraphics[width=1\columnwidth]{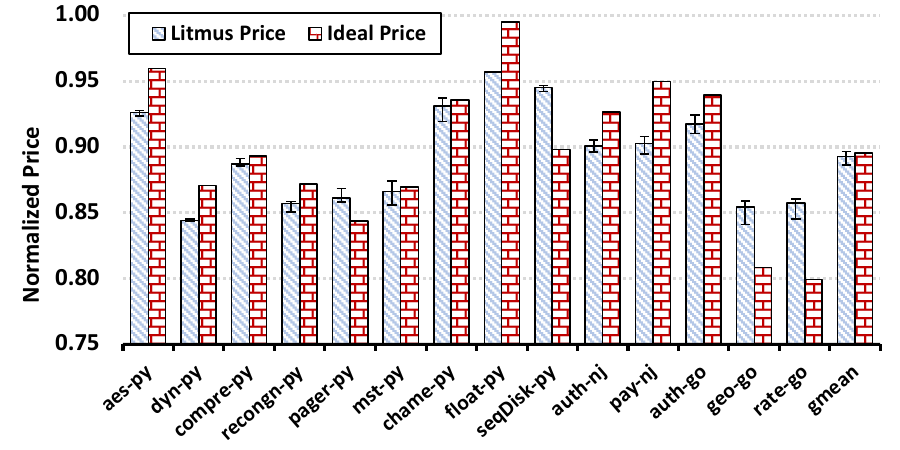}\\
\caption{Prices computed via Litmus pricing and ideal prices when co-running with 26 other functions, normalized to the commercial price that offers no discount}
\label{fig:normalizedPrice26Rev}
\end{figure}


Figure~\ref{fig:normalizedPrice26Rev} illustrates estimated prices derived from Litmus pricing and ideal prices; both are normalized to the price from commercial pricing. The functions receive discounted prices to offset delays in a congested execution environment. The figure shows minimal disparities between the two prices, implying that Litmus pricing offers a reasonable discount. The average discount from Litmus pricing across all functions is 10.7\%, which is only marginally higher by 0.4\% compared to the average slowdown experienced by functions, which stands at 10.3\%, as depicted by the ideal price.

\begin{figure}
\centering
\includegraphics[width=1\columnwidth]{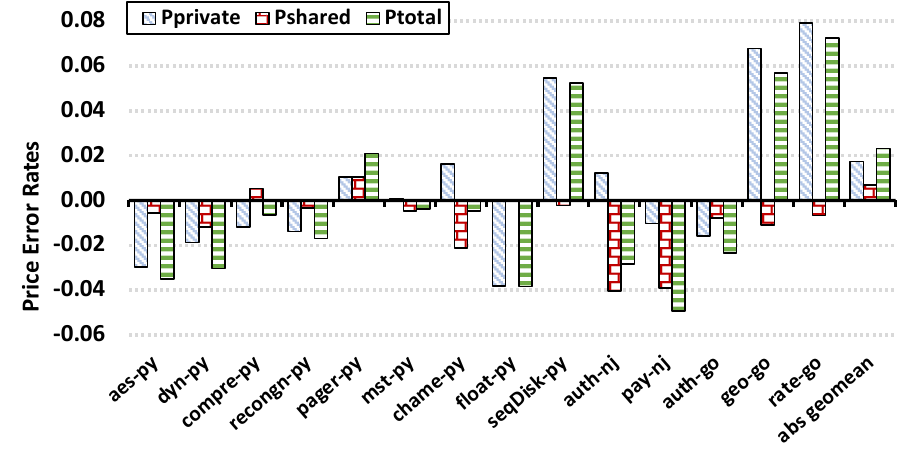}\\
\caption{Weighted errors in prices derived from Litmus pricing, which are compared to ideal prices}
\label{fig:priceErrorRunwith26Rev}
\end{figure}

However, the discounts per slowdown differ among functions. For instance, float-py's ideal discount rate is 0.05\%, but it benefits from a 4.3\% discount with Litmus pricing. Conversely, pager-py encounters a 15.6\% performance decline while receiving only a 13.9\% discount. As shown in Figure~\ref{fig:TpriTshardDistriRev}, pager-py relies more on shared resources with a longer \(T_{shared}\), implying that pager-py's execution time is more susceptible to congestion than the references. In contrast, float-py predominantly utilizes private resources, resulting in minimal impact from the execution environment. Further insights into these variations in each pricing component are presented in Figure~\ref{fig:priceErrorRunwith26Rev}.




Figure~\ref{fig:priceErrorRunwith26Rev} depicts the weighted errors each function experiences. The error rate is computed by comparing the estimated prices with the ideal prices and dividing the difference by the ideal prices. The first two bars show errors in \(P_{private}\) and \(P_{shared}\), weighted by their corresponding ratios in execution time, reflecting their impacts as shown in Figure~\ref{fig:TpriTshardDistriRev}. The last bar shows the error in their total execution time. Functions may have two types of errors: positive and negative. A positive error means that the estimated price exceeds its ideal price, implying that the compensation received is less than the performance loss incurred. In other words, Litmus pricing underestimates the slowdown relative to the actual slowdown. Conversely, a negative error indicates the opposite; a function's compensation exceeds its slowdown. This discrepancy indicates Litmus pricing may offer varying advantages to different functions, favoring those with lesser slowdowns. Meanwhile, tenants must recognize the consequences if their functions differ vastly from the references, particularly those heavily reliant on shared resources.

The largest absolute error is 0.072, occurring with rate-go, while the smallest absolute error is 0.004 with mst-py. Overall, Litmus pricing exhibits minor errors, averaging 0.023. Thus, the estimated discount deviates by 22.5\% from the ideal discount. Note that Litmus pricing aims to match the average discount across all functions by applying a system-wide discount based on the system's congestion level rather than adjusting the discount for each function individually. Consequently, individual functions may experience either positive or negative errors. However, it is noteworthy that the average discount between Litmus and ideal pricing differs by just 0.4\%, as illustrated in Figure~\ref{fig:normalizedPrice26Rev}, satisfying its objectives.
Specifically, Litmus pricing estimates slowdowns in \(P_{private}\) with an average weighted error rate of 0.018. The highest for \(P_{private}\) is 0.079, while the lowest is just 0.001. For \(P_{shared}\), the average weighted error rate is even lower, at 0.007. The highest for \(P_{shared}\) is 0.040.


Litmus pricing estimates each pricing component separately with an individual discount rate. For functions that demand minimal shared resources, such as float-py, the errors in P are nearly identical to those in \(T_{private}\), and the errors in \(T_{shared}\) can be overlooked. As shown in Figure~\ref{fig:TpriTshardDistriRev}, \(T_{private}\) dominates the execution time across many applications. While even a small error in the estimated \(T_{private}\) influences the function's execution time, substantial errors in \(T_{shared}\) likely have limited impacts, except for functions that have a relatively meaningful \(T_{shared}\) ratio, such as pager-py and mst-py, whose errors in \(T_{shared}\) are observable in the total price. 

We emphasize that Litmus pricing cannot avoid producing a certain degree of error since it relies on estimation using a few metrics. However, the figure proves that Litmus pricing effectively captures a congested environment and reflects that environment in the price. Under the same congestion level, \(T_{private}\) values of all functions are delayed similarly with a little variation. Thus, the Litmus pricing can precisely estimate the \(P_{private}\) with the Litmus test, leading to reliable total price estimates since \(T_{private}\) generally dominates execution time. \(T_{shared}\) fluctuates more dynamically across functions and has a relatively larger error, but the impact of this error is minimal, as shown in Figure~\ref{fig:priceErrorRunwith26Rev}.

\begin{figure}[h]
\centering
 \includegraphics[width=1\columnwidth]{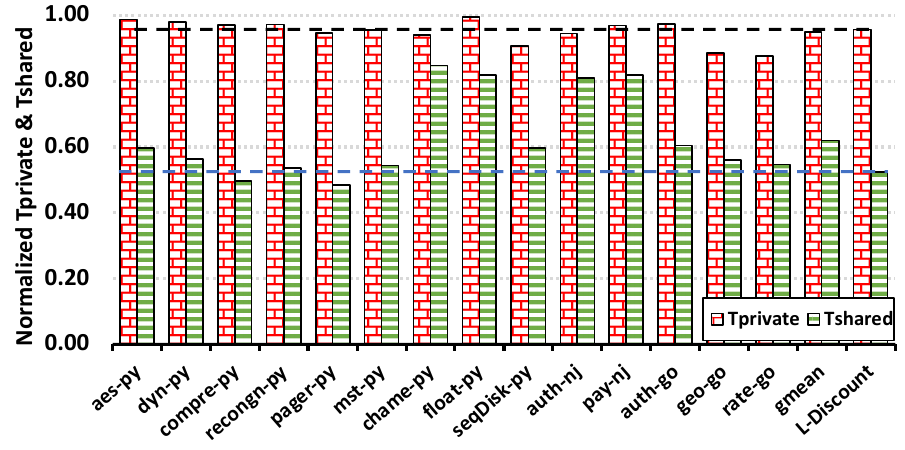}\\
\caption{\(T_{private}\) \& \(T_{shared}\) of applications when co-running with 26 others, normalized to those when they run alone. The top(black) and bottom(blue) lines are discount rates from Litmus pricing for \(T_{private}\) and \(T_{shared}\)}
\label{fig:normTprivtTshared26Rev}
\end{figure}

Figure~\ref{fig:normTprivtTshared26Rev} shows the \(T_{private}\) and \(T_{shared}\) of each function, normalized to the same time components of the baseline. The gap between the bar and 1 represents the amount of slowdown that should be translated into ideal discounts. The top (black) and bottom (blue) dotted lines, along with the last bars, represent the estimated slowdowns of \(T_{private}\) and \(T_{shared}\) with Litmus tests. The figure shows that interferences in private resources extend the function execution time by only 5.3\%, implying the performance of the private resources is marginally affected by congestion. Additionally, all functions exhibit similar \(T_{private}\) values with little dispersion, helping precise estimation with Litmus tests.

Meanwhile, shared resources are more significantly affected than what the Litmus test estimates. However, the errors remain within an acceptable range as their impacts are minor. Functions that have a considerable \(T_{shared}\) ratio are more sensitive to system congestion. Conversely, a function like float-py, which has a negligible \(T_{shared}\) ratio, gains huge benefits from high discounts while maintaining stable performance under congestion. This shows that Litmus pricing accurately reflects the dynamic values of shared resources when the congestion level changes and incentivizes functions to be developed to use less shared resources.

\subsection{Temporal CPU Sharing}
\label{sec:temporalCPUSharing}

The previous evaluation assumes an isolated environment where a CPU is assigned exclusively to an active function during execution, disallowing temporal sharing between functions through context-switching. While this setup ensures strict performance isolation, it inherently limits resource utilization. Conversely, enhancing utilization by permitting temporal CPU sharing among functions complicates price estimation. We explore how Litmus pricing can be applied in a less restrictive environment that permits sharing, where functions can temporarily share the same CPU and are not bound exclusively to specific cores. 

In this section, we deploy a testing function to co-run with 160 other functions across 16 cores without exclusive core assignments, assuming an average of 10 functions share a core. The reduction in the number of cores tested was necessitated by our system's memory limitations, as certain functions require significant memory space.

\textbf{Method 1, Modeling Switching Overhead:} When a function is switched out, a subsequent function evicts the cached data of the displaced one. We noticed that this overhead grows as the number of co-running functions increases. Consequently, permitting more concurrent functions results in slower function execution. However, delayed execution due to extra sharing financially benefits the providers. Thus, we argue that the extent of temporal sharing should be incorporated into Litmus pricing as another discount factor.

\begin{figure}
\centering
\includegraphics[width=1\columnwidth]{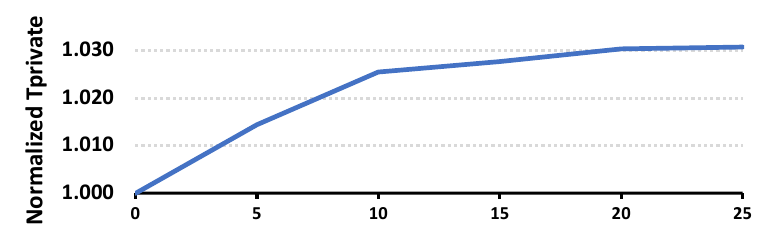}\\
\caption{\(T_{private}\) of functions over different co-running counts in the same core, normalized to when running alone}
\label{fig:csImpactor}
\end{figure}

Figure~\ref{fig:csImpactor} depicts the switching overheads as concurrently running functions increase. The figure demonstrates that the sharing overhead follows a logarithmic growth pattern and stabilizes at around 20 co-running functions. Furthermore, we discovered that the switching overheads predominantly affect \(T_{private}\). Consequently, Litmus pricing needs to calibrate \(T_{private}\) to account for the impacts of sharing before estimating the slowdown using the performance table.

\begin{figure}
\centering
\includegraphics[width=1\columnwidth]{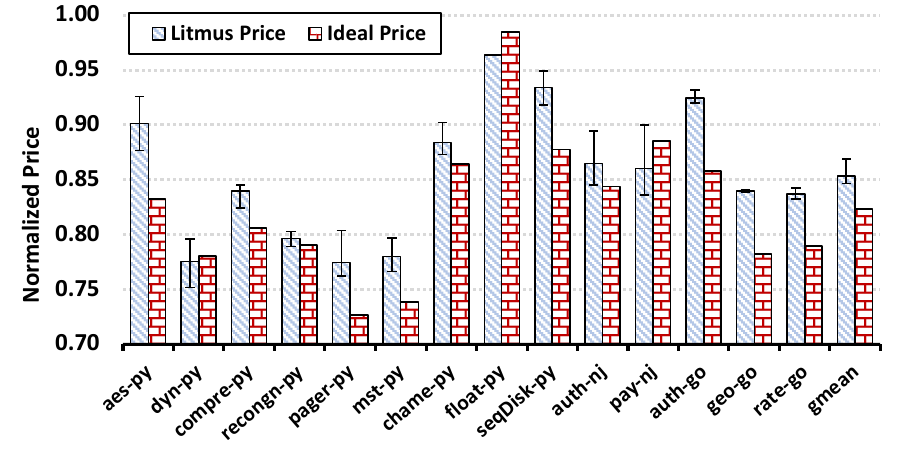}\\
\caption{Prices derived from Litmus pricing (Method 1) and ideal prices when co-running with 160 others, both normalized to commercial prices that offer no discount}
\label{fig:normalizedPrice160Method1}
\end{figure}

Figure~\ref{fig:normalizedPrice160Method1} illustrates the discounted price reflecting this adjustment. In our configuration, with an average of 10 functions per core, we divide \(T_{private}\) by 1.025 before estimating the slowdown. In this figure, Litmus pricing estimates an average discount of 14.5\%, which falls 2.9\% short of the ideal discount of 17.4\%. While most benchmarks receive discounts lower than ideal, aes-py stands out with the largest error of 9.9\%, which is 6.9\% less than its ideal discount of 16.8\%.

\textbf{Method 2, Updating Performance Tables:} Rather than reusing the performance/congestion tables designed for restrictive environments, arranging new tables for the sharing-enabled environments returns better accuracy. Although the notion of preparing separate tables for various sharing levels may seem challenging, we argue that only one or a few extra tables for heavily crowded systems would suffice, given that commercial systems are typically heavily crowded. Again, we emphasize that the impact of function co-placement stabilizes above a certain co-placement count, as shown in Figure~\ref{fig:csImpactor}.

For this test, we prepared new performance/congestion tables for the environment described above, where a testing function is assumed to co-locate and compete for a core with 9 other functions. Moreover, considering the scenario where a switched-out function has a low chance of being rescheduled to the same core, instead of assigning 10 functions to a specific core, we ran 50 functions across 5 dedicated cores; each can run on any of the 5 cores. We managed the congestion levels in shared resources using traffic generators on the other cores, as outlined in Section~\ref{sec:designDetails}.

\begin{figure}[h]
\centering
\includegraphics[width=1\columnwidth]{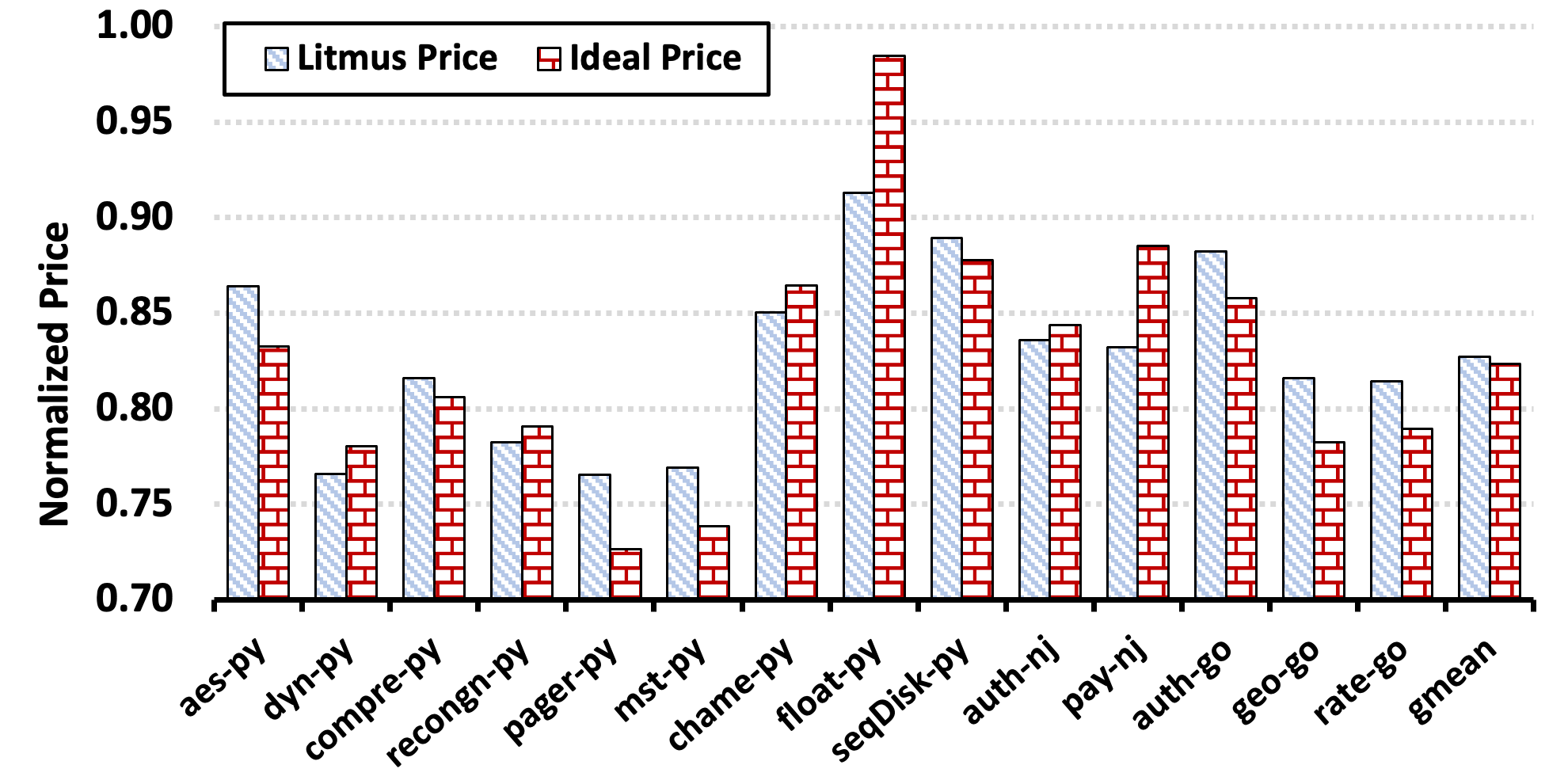}\\
\caption{Prices derived from Litmus pricing (Method 2) and ideal prices when co-running with 160 others, both normalized to commercial prices that offer no discount}
\label{fig:normalizedPrice160}
\end{figure}

Figure~\ref{fig:normalizedPrice160} shows the prices obtained from Litmus pricing with the new tables, which are normalized to those from commercial sources. Once more, as in Figure~\ref{fig:normalizedPrice26Rev}, the figure reveals marginal distinctions between the two price points, suggesting that Litmus pricing continues to offer a reasonable discount. Across all functions, the average discount provided by Litmus pricing stands at 17.2\%, merely 0.2\% less than the ideal discount of 17.4\%.

\section{Sensitivity}
\label{sec:sensitivity}

While the aforementioned tests highlight Litmus's potential, we recognize the need for more compelling results. This section expands our analysis and examines Litmus pricing from various angles across diverse system environments.

\begin{figure}[t]
\centering
\includegraphics[width=1\columnwidth]{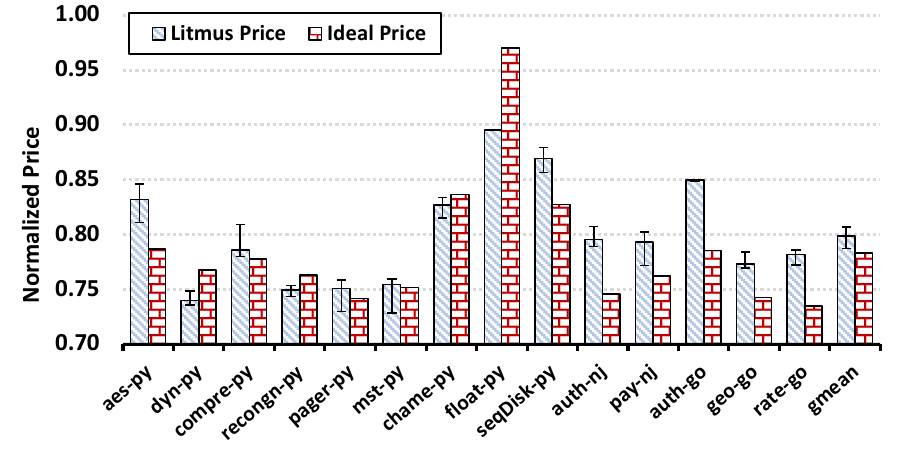}\\
\caption{Prices derived from Litmus pricing (Method 2) and ideal prices when co-running with 320 others, both normalized to commercial prices that offer no discount}
\label{fig:normalizedPriceCongested320}
\end{figure}

\textbf{Heavy Congestion:} We significantly escalate the congestion level to verify Litmus pricing under conditions of substantial slowdowns. Figure~\ref{fig:normalizedPriceCongested320} presents the outcomes of the same tests depicted in Figure~\ref{fig:normalizedPrice160} but with 320 co-running functions. Not only did we increase the function count, but we also specifically selected 8 memory-intensive functions, aes-py, compre-py, thum-py, bfs-py, auth-py, fib-go, geo-go, and profile-go, that produce the most L2 cache misses among the tested functions to create heavy congestion in shared resources deliberately. Even under such intensified congestion, Litmus pricing produces results remarkably close to the ideal price, discounting the price by 20.0\%, which deviates by merely 1.5\% from the ideal discount of 21.5\%. The highest discount by Litmus pricing is 26.0\% for dyn-py, a minor error of 2.8\% compared to the ideal discount.



\textbf{CPU Frequency:} Modern CPUs feature dynamic CPU frequency adjustments, which are controlled via software or hardware. A hardware technique like Intel Turbo Technology autonomously adjusts the CPU frequency based on the CPU's power and thermal budgets, allowing the CPU to operate at higher frequencies when conditions permit and lower frequencies under heavy load. Prior tests assumed that the frequencies were fixed and strictly managed by service providers who had knowledge of when and to what extent frequency adjustments were necessary. Also, we evaluated Litmus pricing under the other scenarios where CPU frequencies were not fixed, providing more dynamic perspectives.

\begin{figure}[t]
\centering
\includegraphics[width=1\columnwidth]{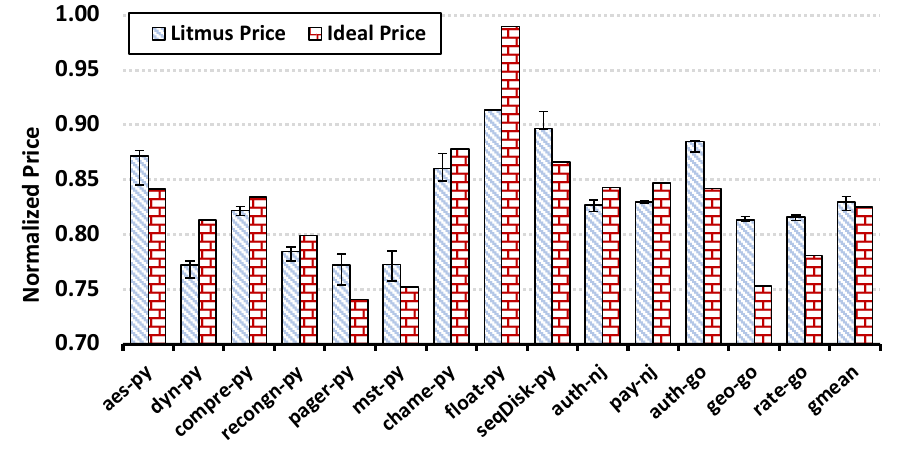}\\
\caption{Prices derived from Litmus pricing (Method 2) and ideal prices when co-running with 160 others with unfixed CPU frequencies, both normalized to commercial prices}
\label{fig:normalizedPrice160unset}
\end{figure}

Figure~\ref{fig:normalizedPrice160unset} illustrates the prices derived from Litmus pricing alongside the ideal prices, both normalized to commercial prices. This evaluation was conducted under the configuration outlined in Section~\ref{sec:temporalCPUSharing} with 160 other functions. Compared with Figure~\ref{fig:normalizedPrice160}, we observe a slight decrease in Litmus pricing's discount, from 17.2\% to 16.8\%, while the ideal discount rate decreases from 17.4\% to 17.3\%. Still, even without fixed CPU frequencies, the difference in the discount rate from the ideal price is only 0.5\%. We noted that CPU frequency changes were infrequent when the system operated with 160 functions. Overall, the variation in frequency had a negligible impact on Litmus pricing.

\textbf{CPU Architecture:} Ensuring the validity of Litmus pricing across different architectures is crucial. Our Litmus tests rely on the performance counter that measures a stall count due to L2 cache misses (cycle\_activity.stalls\_L2\_miss)~\cite{perfMonEvents}, supported by Intel's CPUs. Unfortunately, other vendors like AMD do not yet offer the same performance counter~\cite{amdPMC}, restricting our tests to Intel CPUs. Alternatively, we conducted tests on another Intel CPU based on the Ice Lake architecture, Xeon Silver 4314, to broaden the scope of our analysis.

\begin{figure}[b]
\centering
\includegraphics[width=1\columnwidth]{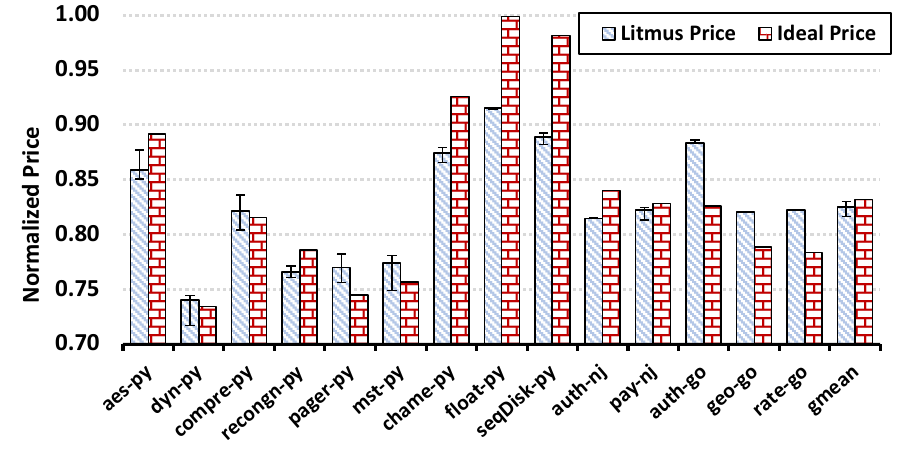}\\
\caption{Prices derived from Litmus pricing (Method 2) and ideal prices when co-running with 70 others on Xeon Silver 4314, both normalized to commercial prices}
\label{fig:normalizedPriceCorun70kiwi}
\end{figure}

Once again, this evaluation was conducted following the configuration outlined in Section~\ref{sec:temporalCPUSharing}, albeit with 70 co-running functions, limited by the main memory capacity of 128GB. Following Method 2, we constructed new congestion and performance tables with 50 functions running across 5 cores. Then, we ran 70 functions across 7 cores to match the competition count, averaging 10 functions per core. Figure~\ref{fig:normalizedPriceCorun70kiwi} presents the results. On average, with Litmus pricing, the tenant only pays 82.5\% of the commercial prices, which is merely 0.7\% less than the ideal price.

\begin{figure}
\centering
\includegraphics[width=1\columnwidth]{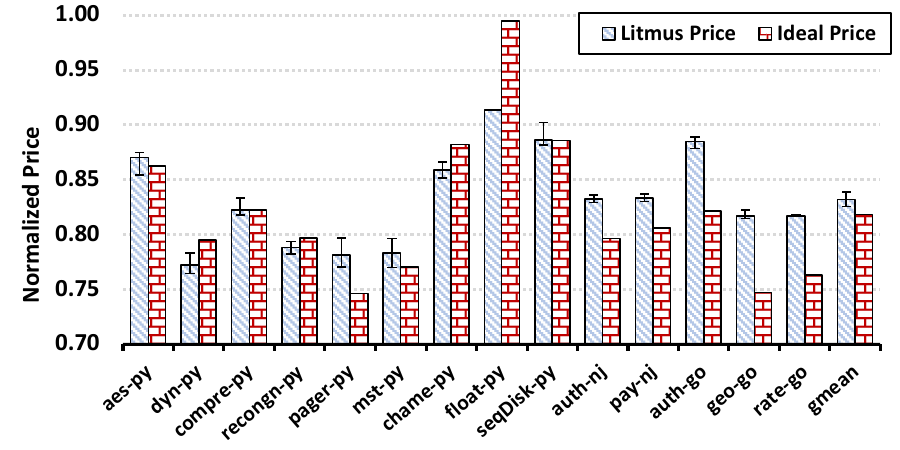}\\
\caption{Prices derived from Litmus pricing (Method 2) and ideal prices when co-running with 15 functions per core and reusing tables, both normalized to commercial prices}
\label{fig:normalizedPriceCorun15}
\end{figure}

\textbf{CPU Sharing Overhead:} The co-located function count determines the level of interference, impacting a function's performance differently. To assess Litmus pricing under various co-running function counts, we increased the function count to 240 from the configuration outlined in Section~\ref{sec:temporalCPUSharing}, making an average of 15 functions running on each core. However, we reused the performance and congestion tables generated in Section~\ref{sec:temporalCPUSharing}. Figure~\ref{fig:normalizedPriceCorun15} illustrates the results, where the Litmus pricing's error is 1.2\% with an average discount of 16.7\%, compared to the ideal of 17.9\%. Despite reusing the tables constructed for 10 co-running functions per core, the error remains negligible. This outcome aligns with Figure~\ref{fig:csImpactor}, which highlights the diminishing impact of temporal sharing when running more than 10 co-running functions. Given the perpetual overcrowding of commercial systems, any configuration gap between the environment where constructing the tables and where conducting tests is expected to remain minor and easily manageable.

\begin{figure}
\centering
\includegraphics[width=1\columnwidth]{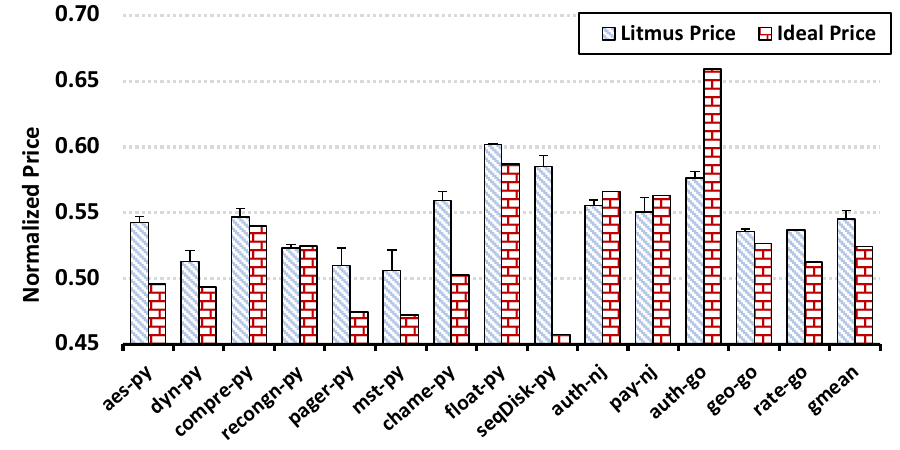}\\
\caption{Prices derived from Litmus pricing (Method 2) and ideal prices when co-running with 160 others in an SMT-enabled system, both normalized to commercial prices}
\label{fig:normalizedPriceCorun160smt}
\end{figure}

\textbf{Simultaneous Multithreading (SMT):} SMT is a technique aimed at maximizing resource utilization by enabling multiple threads to execute concurrently on the same core, extending the shared resource domain across the entire processor. However, while SMT enhances resource utilization, it significantly complicates the measurement of interference impact. Moreover, SMT introduces additional side channels, thereby increasing the processor's vulnerability to security attacks. Due to these concerns, serverless platforms like Amazon Lambda disable SMT in their systems~\cite{schall2023warming,agache2020firecracker,schall2022lukewarm}. Nonetheless, we have conducted a study to assess the impact of SMT on our Litmus pricing methodology.


To construct the performance and congestion tables, we executed 50 applications across 5 physically separated cores, allowing an average of 10 functions to share two virtual cores. Figure~\ref{fig:normalizedPriceCorun160smt} illustrates the results,  presenting prices derived from Litmus pricing and ideal prices, both normalized to commercial prices without discounts. The normalized price experiences a significant drop compared to other configurations, highlighting the impact of heavy congestion within a physical core. The ideal price, which assumes no interference, stands at 47.3\% of the commercial price. Meanwhile, Litmus pricing offers a discount of 45.4\%, which is only 1.9\% less, confirming the effectiveness of Litmus pricing.




\section{Related Work}
\label{sec:related}

Cloud providers strive to maximize profits by accommodating more tenants on their machines, which inevitably introduces interference between tenants, leading to unpredictable application slowdowns. To address this challenge, prior researchers have focused on providing tenants with a more unbiased and predictable quality of service~\cite{kannanproctor,zhao2021understanding,lee2009mcc,jin2015hardware,kannan2019caliper,vicent2017AppClustering,lavanya2015AppSlowdown}. Vicent et al.~\cite{vicent2017AppClustering} proposed clustering-based cache partitioning to mitigate unfairness between tenants. Rohan et al.~\cite{roy2021satori} aimed for both fairness and throughput by simultaneously controlling multiple architectural resources. However, achieving fairness comes at the cost of sacrificing resource utilization. In contrast, Alex et al.~\cite{alex2013ScFairPrice,tiwari2013enabling} addressed the unfairness by adjusting their prices while accepting a certain degree of unfairness. Our study follows a similar direction but explores a more practical solution within the context of serverless computing.

\section{Conclusion}
\label{sec:conclusion}

Serverless computing is a key technology in contemporary cloud computing, offering a range of benefits. One primary advantage is effective cost-saving, as tenants are billed only for the resources they use. However, the time-based fees on commercial platforms can unfairly charge tenants during periods of high congestion, which not only results in slowdowns but also higher costs for tenants. Rather than aiming to maintain service quality, this paper suggests discounting tenants' costs to compensate for performance losses. Litmus pricing proposed in this work monitors machine states through Litmus tests and adjusts tenant costs accordingly. The Litmus test is a lightweight testing approach for serverless platforms, which assesses the machine's congestion level before starting a user's function. Our tests prove that Litmus pricing offers nearly ideal prices in heavily congested environments, with an average deviation of just 0.2\% from the ideal price that adjusts discounts in proportion to the slowdown experienced.

\section{Acknowledgment}
\label{ack}

We would like to thank our reviewers for their valuable feedback to improve this paper. This work was supported by NSF CAREER Award CCF-2146475.


\bibliographystyle{plain}
\bibliography{refs}

\end{document}